\documentclass[12pt,preprint]{aastex}

\usepackage{rotating,lscape,natbib,longtable,times}
\newcommand{\xmm}{{\it XMM-Newton}}
\newcommand{\rosat}{{\it ROSAT}}

\newcommand{\cxo}{{\it Chandra}}

\newcommand{\spitzer}{{\it Spitzer}}
\newcommand{\etal}{et al.}
\newcommand{\NH}{\mbox {$N_{\rm H}$}}
\newcommand{\nh}{\mbox {$N_{\rm H}$}}

\newcommand{\ha}{H{\sc $\alpha$}}

\newcommand{\hi}{H\,{\sc i}}
\newcommand{\hii}{H\,{\sc ii}}

\newcommand{\oiii}{[O\,{\sc iii}]}

\newcommand{\oviii}{O\,{\sc viii}}
\newcommand{\neix}{Ne\,{\sc ix}}

\newcommand{\fel}{Fe\,{\sc l}}

\newcommand{\about}{$\sim$\kern.03em}
\newcommand{\ergs}{erg s$^{-1}$}

\newcommand{\chase}{ChASeM33}
\newcommand{\m}{M33}

\slugcomment{Draft \today}

\shorttitle{\chase: IC131}
\shortauthors{T\"ullmann et al.}

\begin{document}


\title{\cxo\ ACIS Survey of \m\ (\chase): \\ The enigmatic X-ray emission from IC131}


\author{Ralph T\"ullmann\altaffilmark{1},
Knox S. Long\altaffilmark{2},
Thomas G. Pannuti\altaffilmark{3},
P. Frank Winkler\altaffilmark{4},
Paul P. Plucinsky\altaffilmark{1},
Terrance J. Gaetz\altaffilmark{1},
Ben Williams\altaffilmark{5},
Kip D. Kuntz\altaffilmark{6,7},
Wolfgang Pietsch\altaffilmark{8},
William P. Blair\altaffilmark{6},
Frank Haberl\altaffilmark{8},
Randall K. Smith\altaffilmark{1}}
\altaffiltext{1}{Harvard-Smithsonian Center for Astrophysics, 60 Garden Street, Cambridge, MA 02138}
\altaffiltext{2}{Space Telescope Science Institute, 3700 San Martin Drive, Baltimore, MD 21218}
\altaffiltext{3}{Space Science Center, 235 Martindale Drive, Morehead State University, Morehead, KY 40351}
\altaffiltext{4}{Dep. of Physics, Middlebury College, Middlebury, VT 05753}
\altaffiltext{5}{Dep. of Astronomy, University of Washington, Box 351580, Seattle, WA 98195}
\altaffiltext{6}{Dep. of Physics and Astronomy, Johns Hopkins University, 3400 North Charles Street, Baltimore, MD 21218}
\altaffiltext{7}{Astrophysics Science Division, NASA/GSFC, Greenbelt, MD}
\altaffiltext{8}{Max-Planck Institut f\"ur extraterrestrische Physik, Giessenbachstrasse, 85748 Garching, Germany}

\begin{abstract}
We present the first X-ray analysis of the diffuse hot ionized gas and the point sources in IC131, after NGC604 the second most X-ray luminous giant \hii\ region in \m. The X-ray emission is detected only in the south eastern part of IC131 (named IC131-se) and is limited to an elliptical region of $\sim$200 pc in extent. This region appears to be confined towards the west by a hemispherical shell of warm ionized gas and only fills about half that volume. 
Although the corresponding X-ray spectrum has 1215 counts, it cannot conclusively be told whether the extended X-ray emission is thermal, non-thermal, or a combination of both.
A thermal plasma model of $kT_e$\,=\,4.3\,keV or a single power law of $\Gamma$\,$\simeq$\,2.1 fit the spectrum equally well. If the spectrum is purely thermal (non-thermal), the total unabsorbed X-ray luminosity in the 0.35\,--\,8\,keV energy band amounts to $L_X=6.8\ (8.7)\times 10^{35}$\,\ergs. Among other known \hii\ regions IC131-se seems to be extreme regarding the combination of its large extent of the X-ray plasma, the lack of massive O stars, its unusually high electron temperature (if thermal), and the large fraction of $L_X$ emitted above 2\,keV ($\sim$40\,--53\%). 
A thermal plasma of $\sim$4\,keV poses serious challenges to theoretical models, as it is not clear how high electron temperatures can be produced in \hii\ regions in view of mass-proportional and collisionless heating. If the gas is non-thermal or has non-thermal contributions, synchrotron emission would clearly dominate over inverse Compton emission. It is not clear if the same mechanisms which create non-thermal X-rays or accelerate CRs in SNRs can be applied to much larger scales of 200\,pc. In both cases the existing theoretical models for giant \hii\ regions and superbubbles do not explain the hardness and extent of the X-ray emission in IC131-se. We also detect a variable source candidate in IC131. It seems that this object is a high mass X-ray binary whose optical counterpart is a B2-type star with a mass of $\sim$9$M_{\odot}$.
\end{abstract}


\keywords{ISM: \hii-regions --- ISM: superbubbles --- galaxies: individual
(\m) --- X-rays: individual (IC131) --- X-rays: ISM}


\section{Introduction}
This study is part of the \cxo\ ACIS Survey of M33 \citep[\chase,][]{plu08} and presents the first deep, high resolution X-ray images of \object{IC131}. Within \chase's sensitivity limit of $\sim$$2\times 10^{34}$\, erg\,s$^{-1}$ (0.35\,--\,8\,keV) IC131 is, along with \object{NGC604} \citep{tull08}, the only giant \hii\ region (GHR) in \object{\m}\ which shows significant diffuse extended X-ray emission. 

Because of their large sizes, masses, stellar contents, and luminosities, GHRs \citep[cf.][]{ken84} provide excellent laboratories for studying the relation between stellar feedback and the evolution of the individual components of the ISM inside them. GHRs like 30\,Dor \citep{wang99,town06}, N51D \citep{coop04}, or NGC604 \citep{tull08} are strong X-ray emitters thanks to their massive O star population. The heavy mass loss of these stars can produce strong colliding winds which, together with contributions from SNe and SNRs, shock-heat the gas and forces it to emit thermal X-rays at temperatures around $kT_e=0.6\pm0.2$\,keV. 
In some cases the X-ray spectra of \hii\ regions in the Milky Way and the LMC require, in addition to a thermal component, also a non-thermal component \citep[e.g.,][]{wolk02,bam04,coop04,smith04,muno06,mad09}. The non-thermal radiation could be explained by synchrotron emission, inverse Compton scattering, or by particle acceleration in shock regions \citep{par04,arsch08}. In all cases, the X-ray emission could be linked to massive stars or their successors. 
The issue one faces with GHRs is to find a mechanism which can ionize the gas on spatial scales which can easily exceed that of normal \hii\ regions by one to two orders of magnitude and is consistent with the stellar energy feedback, X-ray luminosity, gas mass, and other predictable quantities.

Although IC131 has been observed in numerous M33 surveys across almost all wavelengths \citep[e.g.,][]{rosa84,land92,long96,hippe03,pietsch04,mass06,tab07,plu08}, little is known about the stellar population, its age, and the different components of the ISM. The \chase\ data set, supplemented by other observations at optical and infrared wavelengths, is the first data set that has the combination of S/N and spatial resolution to permit a detailed study of the X-ray emitting gas in IC131. Therefore, the primary focus of this work is to study the hot ionized medium (HIM) by constraining basic plasma parameters, to determine the likely ionization mechanism of the hot gas, and to compare properties of the gas to other star forming regions to learn more about the evolution of the gas in GHRs. The second purpose of this study focuses on the analysis of the point source population in IC131.

In the following, we give an overview of the observations used in this study and the data reduction steps that have been applied (Section \ref{sec2}). We then present high resolution (2$\arcsec$) images of the diffuse X-ray emission in different energy bands and compare these to multi-wavelength data to constrain the morphology of IC131 (Section \ref{spec_img}). A spectral analysis is carried out in Section \ref{spec_res} while Section \ref{spec_dis} discusses the possible origin and ionization mechanism of the HIM. Section \ref{point_dis} deals with the analysis of the X-ray emitting point sources detected in IC131 and Section \ref{con} gives a concluding summary of the most important results.

\section{Observations and Data Reduction}\label{sec2}
 We have chosen for our analysis 3 different ACIS-I fields, namely Fields\,3 and 4 from \chase\ (consisting of ObsIDs 6380, 6381, 6382, 6383, and 7226) and one archival field (ObsID 1730). The location of each field is displayed in \citet{plu08}. In the \chase\ observations IC131 is approximately 4\farcm3 off-axis and in ObsID 1730 $\sim$9\arcmin. 
All observations were performed with ACIS-I in `Very Faint' (VFAINT) mode and all data were reprocessed with CIAO\,4.0.1 and CALDB version 3.4.5.
Basic data reduction was carried out according to the steps outlined in \citet{tull08}. The GTI-corrected total exposure time is 422\,ks. 

We constructed exposure-corrected X-ray images from the merged event list in the following energy bands: 0.35\,--\,8.0\,keV (broad), 0.35\,--\,1.1\,keV (soft), 1.1\,--\,2.6\,keV (medium), and 2.6\,--\,8.0\,keV (hard). All images were binned by 4 pixels to a resolution of \about2\arcsec\ (or 7\,pc) to reduce noise fluctuations in the binned pixels. Since IC131 was observed at off-axis angles $>$4$\arcmin$, the radius of the PSF (90\% encircled energy at 1.49\,keV) in our images is between $2\farcs 1$ and $8\arcsec$. Hence, binning to a pixel size of 2\arcsec $\times$\,2\arcsec does not degrade the angular resolution significantly.

In addition to the X-ray data, we also employed archival broadband (R, V) and narrowband imagery (\ha, \oiii $\lambda$5007\AA) from the Local Group Galaxies Survey \citep[LGGS,][]{mass06}, the MMT ($r$, $g$, $u$), and \spitzer\ ({\tt IRAC/MIPS} images at 3.6, 8, and 24$\mu$m) which allow a multi-wavelength comparison of the different phases of the ISM. Because we use these images merely for a morphological comparison, only standard data calibration was applied, except for the \ha\ and \oiii\ images, which have R and V band continuum images subtracted.

We extracted spectra from the diffuse X-ray emission after point source removal. Because the IC131 pointings do not share common roll and pointing directions we did not create spectra from the merged event list. Instead we extracted source and background spectra for each individual event list and merged them by following again the steps described in \citet{tull08}. The resulting source spectra were background subtracted and then grouped using the grouping scheme of ACIS Extract \citep{Broos02} until a signal to noise ratio (S/N) of 3 per channel was achieved and fitted with {\tt XSPEC}\,{\tt ver.\,12.4.0} \citep{Arn96}. Uncertainties are given on a 90\% confidence level and were calculated with the {\tt error} routine. 

\begin{figure*}[pht]
\centering
\includegraphics[width=16cm,height=20.5cm,angle=0]{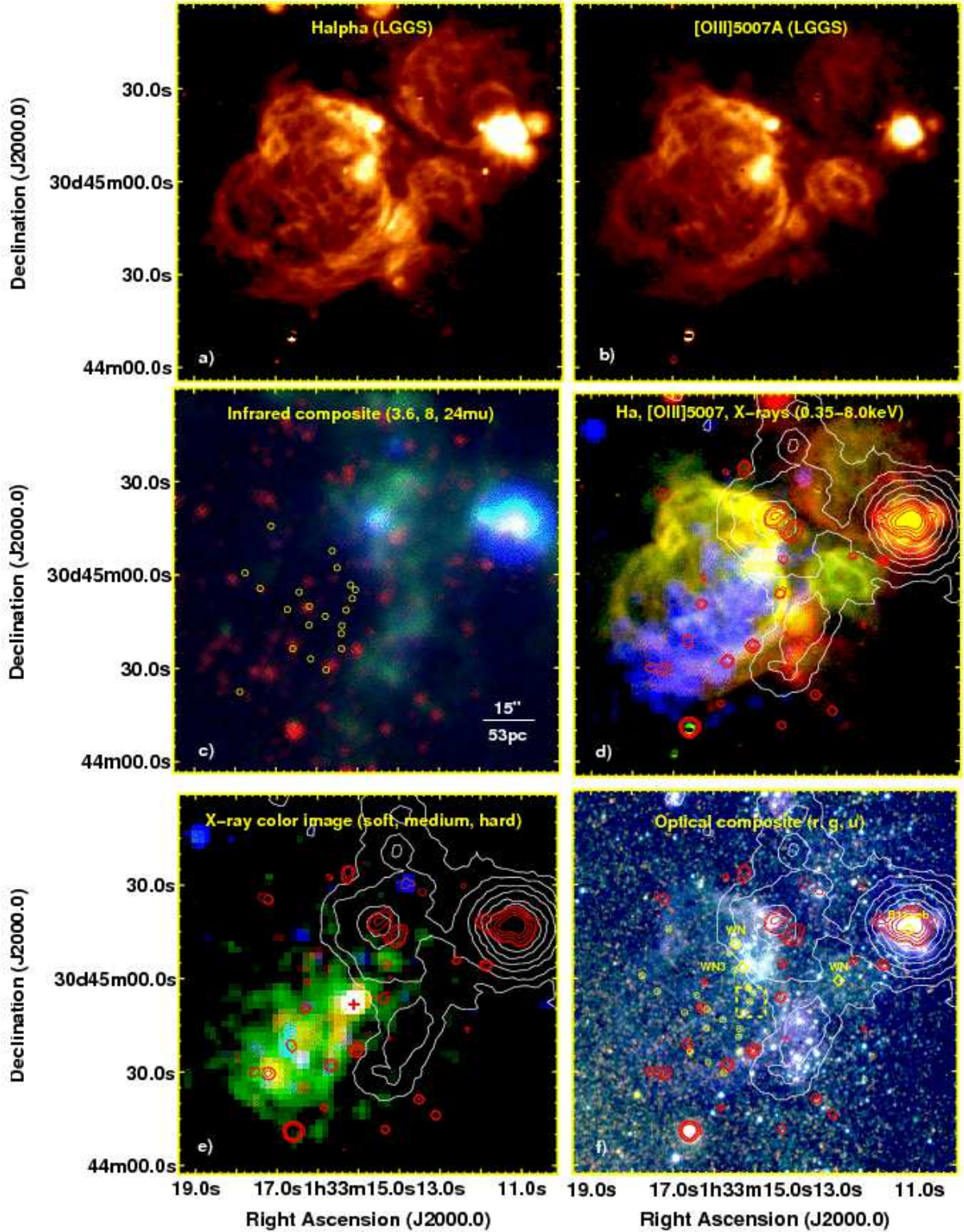}
\caption{\label{f1} {\small (a): Continuum-subtracted \ha\ image of IC131 from the LGGS. (b) Same as a), but for \oiii $\lambda$5007\AA. (c): \spitzer\ {\tt IRAC} and {\tt MIPS} data obtained at 3.6$\mu$m (red), 8$\mu$m (green), and 24$\mu$m (blue).}}
\end{figure*}

The spectrum of the only X-ray detected point source in IC131-se, CXO\,J013315.10+304453.0 \citep[hereafter also FL073 as it is source number 73 in the \chase\ First Look catalog of][]{plu08}, was extracted in a similar manner. This source is located at RA $01^{\rm h}33^{\rm m}15\fs10$ and Dec $30\degr44\arcmin53\farcs00$ (see red cross in Fig.~\ref{f1}e) at the western tip of the extended X-ray emission.

\clearpage
{\small \parindent0pt (d): Multicolor image, showing \ha\ in red, \oiii$\lambda$5007\AA\ in green, and X-rays (0.35\,-8.0\,keV) in blue. (e): High resolution (2\arcsec) Chandra X-ray images (red: 0.35\,--\,1.1\,keV, green: 1.1\,--\,2.6\,keV, blue: 2.6\,--\,8.0~keV). (f): Optical false color composite consisting of MMT broadband $r$, $g$, and $u$ imagery. Positions of three spectroscopically classified WR stars (see labels in panel f) for their spectral type) and 21 unclassified stars within the bubble (yellow circles) are taken from \citet{mass06}. A zoom into the region surrounded by the yellow dashed box is shown in Fig.~\ref{f5} (see Sect. 4.1) and reveals a possible optical counterpart of FL073, the bright X-ray point source marked by a red cross in panel e). Contours in panels d)\,--\,f) represent IR emission at 3.6$\mu$m (red, 0.42\,--\,1.50\,mJy/sr) and 24$\mu$m (white, 31.9\,--\,39\,mJy/sr). See the electronic edition of the Journal for a color version of this figure.}


\section{Diffuse emission}
\subsection{Imaging Results}\label{spec_img}
From the \ha\ and \oiii$\lambda 5007$\AA\ imaging shown in Figs.~\ref{f1}a, b, and, d one can see that IC131 consists of two prominent regions, a hemisphere in the north west and a hemisphere in the south east. The brightest optical emission in the former region seems to be related to an \hii\ region which is ionized by a B1 star \citep{mass06}, whereas the latter complex is the most dominant feature in the optical. This region appears to be a superbubble of ionized gas with a radius of $\sim$100\,pc that opens to the SE (called IC131-se, hereafter). These two complexes appear to be separate bubbles and give IC131 a characteristic dumbbell or hourglass shape.
The western edge of IC131-se shows enhanced \ha\ and \oiii\ emission, running roughly from RA $01^{\rm h}33^{\rm m}15\fs0$ and Dec $30\degr45\arcmin25\farcs0$ to RA $01^{\rm h}33^{\rm m}14\fs0$ and Dec $30\degr44\arcmin30\farcs0$. This area can be associated with young stars, among them are three WR stars, which seem to have formed along the edge (see yellow diamonds in Fig.~\ref{f1}f for their positions and type). It appears likely that these stars photoionize the gas and cause the optical enhancement. Moreover, the IR emission at 24\,$\mu$m (white contours in Figs.~\ref{f1}d\,--\,f), a tracer of warm dust, is well aligned with the western edge of the bubble and seems to follow the stellar distribution. From Figs.~\ref{f1}d and e it can be seen that except in the east, the diffuse X-ray emission seems to be confined by the superbubble and that it extends eastward, away from the loci of the hot stars. 
A peculiarity in the large scale distribution of the X-ray emission in IC131-se is noteworthy. From the image of the 0.35\,--\,8.0\,keV emission shown in Fig.~\ref{f1}d, it appears as if the diffuse gas is confined to the southern half of the superbubble. Such a distribution is unexpected, especially if the gas was produced by stellar winds, SNRs, and SNe, because the high-pressure and supersonically streaming gas should fill the bubble almost completely. Shearing by differential rotation \citep[e.g.,][]{pal90} is an unlikely explanation, as that would equally affect the bubble and the HIM and is likely to act perpendicular to the major axis of the HIM. It would be very illuminating to know how such a lopsided morphology can be produced. 

Using {\tt funcnts} from the FITS utility package {\tt Funtools}\footnote{http://www.cfa.harvard.edu/$\sim$john/funtools/}, which counts the number of photons in a specified regions, we determined that $\sim$85\% of the detected diffuse X-ray photons in IC131-se have energies $>$1\,keV (see the greenish emission in Fig.~\ref{f1}e) and 34\% are emitted above 2\,keV. For NGC604, the second largest \hii\ region in the Local Group, the corresponding values are 45\% and 10\%, respectively, and clearly indicate that the X-ray emission in IC131-se is significantly harder.

Although there are some NIR point sources in IC131-se seen superposed on the X-ray emission, the extended NIR emission and the X-rays are anticorrelated, suggesting that they do not occupy the same volume. This anticorrelation cannot easily be explained in terms of absorption, since most of the X-rays we see have energies in excess of 1\,keV and should not be significantly absorbed. 

Among the 21 identified stars located inside the bubble (yellow circles in Fig.~\ref{f1}), none were spectroscopically classified as type WR or O \citep{mass06}. This implies that most massive stars inside the bubble have already exploded and that we face IC131-se in a relatively late evolutionary stage, i.e. close to the end of the life cycle of massive stars.
The SNe events could have triggered the formation of a sparse, new generation of stars which seemed to have formed along the western edge of the bubble; among them are three WN-type stars \citep{mass06}. This new star formation site is most likely associated with an extended reservoir of molecular gas which could be pristine gas from the remainder of the parent molecular cloud. It appears as if star formation progresses from east to west and has already consumed a large part of the molecular cloud (provided the bubble formed from the same material). 
In this picture, SNe shocks or stellar winds as well as the hot gas itself \citep{smith96} could have swept away or destroyed any significant preexisting accumulation of dust from the bubble's interior, which would also explain the lack of IR emission. 

Apparently, IC131-se is very different from 30\,Dor, NGC604 or any other known superbubble in at least two important aspects: first, IC131-se does not show a clear spatial correlation between young stars and diffuse X-ray emission and second, the X-ray emission in IC131-se (regardless whether thermal or non-thermal) seems to be much harder on spatial scales that have not been observed in other \hii\ regions or superbubbles before.
The first aspect is explainable by different ages of the stellar population, the latter, however, remains puzzling and requires an analysis of possible emission mechanisms.
 
\subsection{Spectroscopic Results}\label{spec_res}
\begin{figure*}[t]
\centering
\includegraphics[width=16.5cm,height=6cm,angle=0]{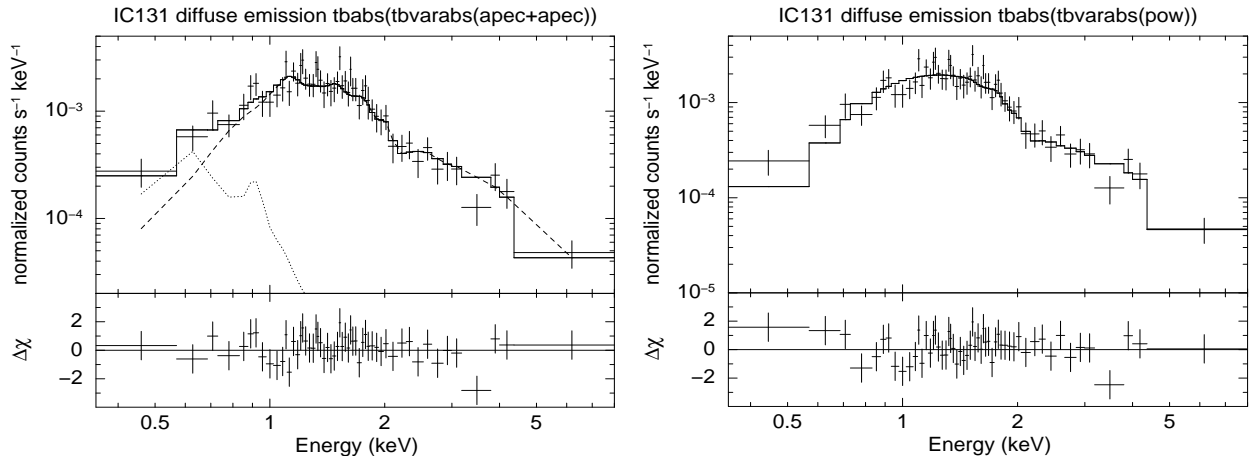}
\caption{\label{f2}
In the left panel the X-ray spectrum of the diffuse emission is fitted with a soft (dotted line) and hard (dashed line) thermal plasma model, whereas the right panel shows the best-fit of a non-thermal power law model.}
\end{figure*}

As discussed in the previous section, the X-ray spectrum of the superbubble in IC131 is much harder than usually observed in other typical GHRs, such as NGC604. 
To explore this, we extracted a spectrum of the diffuse emission, after excluding a region around the point source, and performed spectral fits to help us understand the nature of the X-ray emission.
This spectrum contains 1215 counts and is shown in Fig.~\ref{f2} together with the best-fit models for a purely thermal plasma (albeit two thermal components) and a non-thermal plasma (see also Table~\ref{t1}).
Although the spectrum is relatively smooth, there are some features at soft energies around 0.68 and 0.9\,keV which suggest the presence of \oviii, \neix, or \fel\ emission. At energies above 2\,keV the spectrum looks relatively smooth and there is no real indication of emission lines, such as Fe K\,$\alpha$ at 6.7\,keV. However, it should be kept in mind that $Chandra's$ sensitivity is low at these energies.

\begin{deluxetable}{l|ccc}
\tablecaption{\label{t1} Fit parameters (diffuse emission)}
\startdata
\hline\hline \\[-0.5cm]
tbabs$\times$(tbvarabs$\times$\ldots)\tablenotemark{a}        & (apec)                 & (apec+apec)          & (pow)               \\[0.1cm]
\hline \\[-0.5cm]                                                                                             
$\chi^{2}_{\rm red}$                                            & 0.92                   & 0.86                 & 0.92                \\
degrees of freedom                                            & 54                     & 52                   & 54                   \\
$P$ value\tablenotemark{b}                                    & 0.66                   & 0.75                 & 0.65                 \\[0.1cm]
$K_{1}$ ($10^{-5}\,{\rm cm}^{-5}$)\tablenotemark{c}             & 2.29$^{+0.24}_{-0.15}$   & 5.40$^{+1.93}_{-0.43}$ & 0.92$^{+0.27}_{-0.20}$ \\
$K_{2}$ ($10^{-5}\,{\rm cm^{-5}}$)\tablenotemark{c}             & ---                    & 2.69$^{+1.11}_{-0.41}$ & ---                  \\[0.1cm]
\hline \\[-0.5cm]                                                                                               
$N_{\rm H}$ ($10^{22}\,{\rm cm}^{-2}$)                           & 0.01$^{+0.12}_{-0.01}$  & 0.19$^{+0.49}_{-0.19}$ & 0.17$^{+0.07}_{-0.11}$ \\
$\Gamma$\tablenotemark{d}                                      & ---                   & ---                  & 2.13$^{+0.26}_{-0.19}$ \\
$kT_{1}$ (keV)\tablenotemark{e}                                & 4.32$^{+1.14}_{-0.85}$  & 0.14$^{+0.05}_{-0.10}$ & ---                   \\
$kT_{2}$ (keV)\tablenotemark{e}                                & ---                   & 3.41$^{+1.27}_{-1.24}$ & ---                   \\
$\tilde{F}_{\rm X}$ ($10^{-14}$\,erg/s/cm$^2$)\tablenotemark{f} & 3.05$^{+0.17}_{-0.27}$  & 3.06$^{+0.28}_{-0.67}$ & 2.87$^{+0.24}_{-0.33}$  \\
$F_{\rm X}$ ($10^{-14}$\,erg/s/cm$^2$)\tablenotemark{f}         & 3.39$^{+0.19}_{-0.30}$  & 6.17$^{+0.56}_{-1.35}$ & 4.39$^{+0.37}_{-0.50}$  \\
$L_{\rm 0.35-8}$ ($10^{36}$\,erg/s)\tablenotemark{g}            & 0.68$^{+0.04}_{-0.06}$  & 1.23$^{+0.11}_{-0.27}$ & 0.87$^{+0.07}_{-0.10}$  \\[0.1cm]
$L_{\rm 2-8}$ ($10^{36}$\,erg/s)\tablenotemark{h}               & 0.36$^{+0.03}_{-0.05}$  & 0.35$^{+0.09}_{-0.21}$ & 0.34$^{+0.05}_{-0.08}$  \\[0.1cm]
\enddata
\vspace{-0.5cm}
\tablenotetext{a}{The {\tt tbabs} model assumes a fixed Galactic $N_H$ column density of $6\times10^{20}$\,cm$^{-2}$ \citep{dl90} and solar abundances, whereas the {\tt tbvarabs} model accounts for internal absorption in M33 and uses abundances of $Z=0.6Z_{\odot}$. $^b$Under the assumption that the model is a good fit to the data, the $P$ value is the probability of getting a $\chi^2$ which is greater than the observed $\chi^2_{obs}$. $^c$$K_{1,2}$ are the normalization constants of the fit for the individual model components. $^d$Photon index. $^e$$kT_{1,2}$ represent the electron temperatures of the individual model components. $^f$Absorbed ($\tilde{F}_{\rm X}$) and unabsorbed ($F_{\rm X}$) fluxes are tabulated in the 0.35\,--\,8.0\,keV energy band. $^{g,h}$Unabsorbed luminosities in the 0.35\,--\,8.0\,keV and 2.0\,--\,8.0\,keV energy band, respectively, assume a distance to M33 of $D=817$\,kpc \citep{freed01}.}
\end{deluxetable}
\begin{figure}[!t]
\includegraphics[width=5cm,height=7.5cm,angle=270]{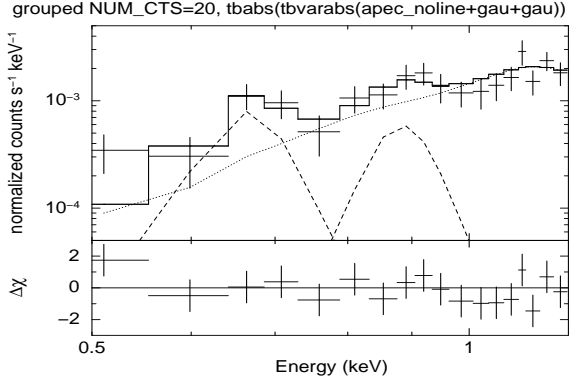}
\caption{\label{f3}
This plot shows part of the diffuse spectrum which was grouped with {\tt dmgroup} until each group had at least 20 counts. It was fitted with an absorbed thermal {\tt apec} model from which the line emission was removed (dotted line) and two Gaussian profiles (dashed lines).}
\end{figure}
\begin{figure}[!t]
\includegraphics[width=5cm,height=7.5cm,angle=270]{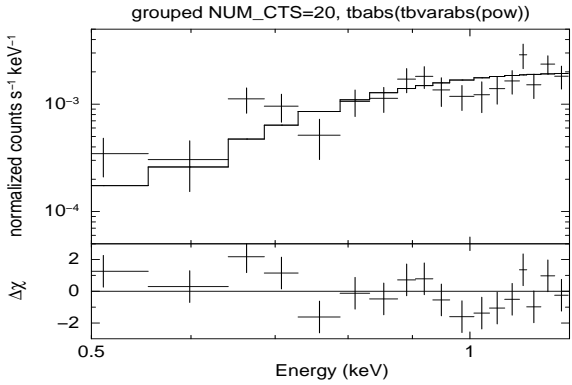}
\caption{\label{f4}
If we fit a simple absorbed power law to the same spectrum shown in Fig.~\ref{f3}, the residuals at 0.68 and 0.89\,keV do not appear significant enough to claim the existence of emission lines, implying that the power law is an acceptable fit, too.}
\end{figure}
By adopting an alternative grouping scheme provided by {\tt CIAO} ({\tt dmgroup}, {\small NUM\_CTS=20}), we tried to increase the number of groups in order to improve the spectral resolution around the energies of the potential emission features (see Fig.~\ref{f3}). 
The resulting spectrum was fitted with a modified {\tt apec} model which has continuum, but no line emission \citep{smith01} and two unconstrained Gaussian profiles at 0.68 and 0.89\,keV. Because this model fits the emission line features well and because the line widths of the Gaussians are small ($\le0.02$\,keV), the presence of a thermal component seems to be indicated.
However, we also fitted the spectrum shown in Fig.~\ref{f3} with an absorbed power law and found that the residuals are barely significant (Fig.~4). Although the present data are suggestive of a thermal component, it is not required.

Since the line features (if they exist) are weak, we also fitted the spectrum with a number of simple models using thermal and non-thermal components. The results of the best model fits are listed in Table~\ref{t1}. In all cases the Galactic \hi\ column density was modeled with the {\tt tbabs} model \citep{wilms00} and fixed at $6\times10^{20}$\,cm$^{-2}$ \citep{dl90}. For the absorption in \m, we used the {\tt tbvarabs} model, adopted a metal abundance of 0.6$Z_{\odot}$ \citep{vil88}, and set the $N_H$ initially to $1\times10^{21}$\,cm$^{-2}$ to account for the average value in \m\ \citep{newton80}. The \nh\ in \m\ was allowed to vary freely during the fitting process. For the {\tt apec} model, the metal abundance was also set to 0.6$Z_{\odot}$.

From a statistical point of view all of the models shown in Table~\ref{t1} are acceptable. An absorbed thermal plasma model appears to be as likely as an absorbed power law. The predicted electron temperature of the thermal model of about 4.3\,keV would be the highest determined so far for any GHRs. However, the unabsorbed luminosity of $\sim$$7.0\times10^{35}$\,\ergs\ is about half that of NGC604.  About 50\% of the absorption-corrected X-ray luminosity of IC131-se is emitted in the 2\,--\,8\,keV energy band, which indicates that the emission is much harder than in ``normal'' giant \hii\ regions, such as NGC604 or 30\,Dor, for which the corresponding fraction is $<$5\% \citep{town06,tull08}. 
As there is no physical reason to believe that the thermal plasma within the complex environment of a GHR would have a single temperature (1-T), we also considered a two-component thermal model (2-T). The best fit has a soft (0.14\,keV) and a hard (3.4\,keV) component and is, not surprisingly, somewhat better than the one from the single temperature model. This second component allows for a significantly better fit of the soft emission below 0.8\,keV and likely yields a more accurate estimate of the X-ray luminosity.
The temperature of the higher temperature component in the 2-T model is lower than the temperature for the 1-T model, which is likely a consequence of the larger $N_H$ of the 2-T model. However, the temperature of the higher temperature component of the 2-T model is consistent (within errors) with the temperature of the 1-T temperature  model.
We also fitted the spectrum with NEI and bremsstrahlung models, which also produced acceptable fits. However, they are not well motivated physically, given that the ionization timescale of about $4\times10^{13}$\,s cm$^{-3}$ as derived from the former model implies ionization equilibrium conditions and the latter considers the unrealistic case of a collisional plasma that is pure H and He.

Non-thermal X-rays, either through synchrotron emission or inverse Compton scattering are well known from SNRs and \hii\ regions. This, together with the fact that the spectrum of IC131-se appears relatively smooth above 2\,keV, motivated us to also fit the spectrum with a purely non-thermal model and a combination of thermal and non-thermal plasma models. Like a single {\tt apec} model, the single power law provides a good fit to the data, too. The photon index of $\Gamma$\,=\,2.13$^{+0.26}_{-0.19}$ is within a reasonable range expected for SNRs and superbubbles with non-thermal X-ray emission and the $N_{\rm H}$ is consistent with the integrated line-of-sight column density.

We also considered combinations of thermal and non-thermal emission, which seemed in some sense natural, given that a combination of thermal and non-thermal emission has also been reported for other GHRs and superbubbles. These models also led to statistically acceptable fits, and the best fit has a photon index of $\Gamma=2.66$ and a very soft plasma temperature of 0.07\,keV. However, the best-fit column density of $\nh=8.4\times 10^{21}$\,cm$^{-2}$ appears unphysical. Balmer line extinction measurements in IC131-se indicate that reddening is low and mostly Galactic \citep{relk09}. 
 Likely as a consequence of the large column density in this model, the unabsorbed luminosity in the broad energy band is unacceptably high, about a factor 60 higher than predicted by the other models and almost a factor of 30 larger than in NGC604. Constraining \nh\ to acceptable values tends to drive the best fit to a simple power law with little or no contribution from the thermal component, which is understandable since the pure power law fits actually do fairly well in fitting the overall shape of the spectrum. We also tried a combination of an {\tt apec} and a {\tt srcut} \citep{rey99} model. This best-fit model predicts a luminosity which is comparable to that of 30\,Dor and a spectral radio index which is much higher than the observed one of $\Gamma =0.3$ (F. Tabatabaei, priv. comm.). Given the limited number of counts in our spectrum and the fact that all of the models we considered are very simple, we do not believe that the combination of thermal and non-thermal emission should be ignored, although the model results do not support that picture.

To ensure that the emission was not contaminated by background sources or insignificant X-ray point-like sources, we also removed the X-ray emission within the cone which coincided with the IRAC $3.6\mu$m emission and refitted the spectrum. No significant differences to the previous best fit results were obtained.
Obviously, one would like to obtain a better X-ray spectrum of IC131-se, but that may be unrealistic for the foreseeable future, because our $Chandra$ exposure totaled 422 ks.
Unfortunately, the available \xmm\ observations provide no further insights as the exposure time is either too low to extract a spectrum, IC131-se falls onto a chip gap or is more than 8\arcmin\ off-axis, where it is strongly vignetted.

\subsection{Discussion}\label{spec_dis}
For a GHR or superbubble the X-ray properties of IC131-se are extreme.
The fraction of the hard X-ray energy flux above 2\,keV exceeds that of 30\, Dor and NGC604 by one order of magnitude (50\% vs. 5\%). If the emission is thermal, the temperature of the hot gas in IC131-se is 4.3\,keV, at least a factor of seven higher compared to these and other known \hii\ regions and superbubbles, such as N11 \citep{mad09}, N51D \citep{coop04}, and 30\,Dor C \citep{bam04,yama09}. This raises the question whether the X-ray emission is a relic of the same process that caused the expansion of the bubble or if it is due to processes which happened in the recent past. In order to answer that and to constrain the nature of the emission, we evaluate the possibilities that the emission is either thermal, non-thermal, or a combination of both. Two other factors need to be incorporated into any answer to this question. IC131-se and the X-ray emitting region are large, a little more than 200\,pc across, and this region contains few if any O stars to provide a continuing source of energy to power the emission.
\subsubsection{Is the X-ray emission thermal?}
There are many reasons to believe that the X-ray emission from IC131-se is due to emission from a thermal plasma. Firstly, the X-ray emission from most GHRs is dominated by emission from thermal plasmas. Secondly, a well-developed explanation for the X-ray emission exists in the form of shocks created by a combination of stellar winds and SNe explosions. Thirdly, the X-ray luminosity of IC131-se is not extreme and is comparable to other superbubbles with thermal emission. Fourthly, from an observational perspective, the single absorbed {\tt apec} model with a $kT_e$ of 4.3\,keV provides an adequate fit to the data, and of all of the thermal models which do so, a low value of \NH\ is predicted that is consistent with other observations. Furthermore, there are hints of line features at positions expected from a thermal plasma. 

The accepted interpretation for X-ray emission from  \hii\ regions and superbubbles \citep[e.g.,][]{weaver77,mk87,mm88,shull95} is that it arises from a thermal plasma that has been heated by supersonic colliding winds from massive stars, SNe, and SNRs to temperatures $\lesssim$1\,keV. 
Within this context the data discussed in Section \ref{spec_img}  suggest that in the case of IC131 a large superbubble formed by the combined action from stellar winds, SNRs, and SNe, filling its interior with a hot X-ray emitting gas, similar to 30\,Dor or NGC\,604. If this interpretation is correct, the absence of a hot and massive stellar population within the bubble implies that these stars already exploded as SNe and the superbubble is a relatively old structure \citep[$>$8\,Myr,][]{maed94}. It might also suggest that the south eastern part of the superbubble was swept away by SNe blast waves. Alternatively, the bubble could have become unstable after the expansion slowed down and cooling set in, which finally led to large-scale breakout of hot gas into the ambient ISM. Either case suggests that IC131-se is in an advanced evolutionary stage. This is different from NGC604, the other X-ray bright GHR in \m, because in this case the present stellar generation in the western part appears to be too young to produce a significant number of SNe.
An evolved and aged superbubble would also explain why no SNRs have been reported for IC131-se, they could have been thermalized a long time ago. It could also be that recent SNe have escaped detection, because they are evolving into a low density medium.

The main difficulty with a thermal interpretation of the emission from IC131 is the high electron temperature. Hot X-ray plasmas ($kT_e$\,$>$\,3\,keV) extending on scales of $\sim$200\,pc have not been reported for GHRs so far. If the plasma is purely thermal, IC131 is the first such region, allowing us to study gas under extreme conditions and to learn more about the mechanism, which can produce such hot plasmas. Before attempting to answer whether the conditions under which the plasma might be heated to these temperatures exist, it is worthwhile to explore whether such a hot plasma is allowed based on energy or mass arguments.

The electron density can be calculated from the normalization constant of the best-fit model (here the single temperature {\tt apec} model) and amounts to $n_{e}=(0.045\pm 0.014)\,f^{-1/2}_{\tiny X}$ cm$^{-3}$ (with $f_{\tiny X}$ being the filling factor). This value is very low (a factor $\sim$5 lower compared to NGC604) and would also support the idea that blowout in IC131-se occurred.

The internal thermal energy within the superbubble can be written as:  
\begin{equation}
E_{th} = 2.88\, n_e\, kT_e\, V f_{\tiny X},
\end{equation}
where $n_e$ is the electron density, $k$ is the Boltzmann constant, $T_e$ is the assumed electron temperature of $T_e=(5.0\pm1.3)\times 10^7$\,K, and V is the volume occupied by the HIM, which we adopt to be $(7.0\pm1.0)\times 10^{61}$\,cm$^{3}$ (an ellipsoid with half-axis radii of 110\,pc\,$\times$\,72\,pc\,$\times$\,72\,pc with 50\% uncertainty in each axis). The internal energy is calculated for a fully ionized plasma of Hydrogen with 10\% Helium and a total number particle density of $n\simeq 1.92n_e$. 
We estimate the internal energy of the bubble to be $(5.9\pm2.9)\times 10^{52}$\,erg, assuming a filling factor of $f_{\tiny X}=0.9$. Because the average conversion efficiency between the kinetic and thermal energy in a SN explosion is unlikely to be one, but closer to 0.5 during the first few 10$^3$\,yr of the evolution of a SNR \citep{tang05}, the thermal energy content in IC131-se should roughly correspond to 118$\pm$58 SNe, assuming each SN releases a kinetic energy of $10^{51}$\,erg. Is the number of SNe also consistent with the number of O-stars expected from the initial mass function (IMF)?

A stellar population analysis of the stars along the line of sight to IC131-se does not currently exist, but the available optical photometry of the stellar population inside the bubble can provide some clues on the nature of the X-ray emission. 
Among the 21 stars seen in projection against the bubble, five appear to be located at the edge of the new star formation site which coincides with the eastern edge of the molecular cloud. Because these stars are possibly not members of the stellar population inside the bubble, we excluded them from the following calculations. 
According to \citet{mass06}, the remaining 16 stars have apparent (absolute) visual magnitudes ranging from 20.13 (-4.65) to 22.99 (-1.78) which implies that their spectral types range from B1\,{\small III} to B6\,{\small III} (7$\le$\,M/M$_{\odot}$\,$\le$\,20) \citep{sk82} and all stars, except one, are potential SN progenitors \citep{woos02}.

These findings confirm our previous consideration that we observe IC131-se during an advanced evolutionary stage in which all O stars already exploded and SNe and SNRs heated the gas. 
Based on the available optical data we can narrow down the age of the stellar population inside the superbubble. Making use of the stellar mass-luminosity relation and assuming that the least massive B star has a mass of $\sim$7\,$M_{\odot}$ and that the lifetime of the least massive O star is about 8\,Myr \citep{maed94}, the age of the stellar population ranges from at least 8\,Myr to 77\,Myr. 
By adopting a Salpeter IMF with slopes ranging from 1.6\,$\le$\,$\alpha$\,$\le$\,3.0 \citep{krou01}, we determine the number of expected O stars with masses 10\,$\le$\,$M/M_{\odot}$\,$\le$\,120 to range from 9 to 21 (it needs to be kept in mind that the normalization constant of the IMF may seriously affect the expected number of O stars if the observed sample suffers from incompleteness). 
This estimate is inconsistent with the number of SNe derived from the internal energy content of IC131-se and argues against a thermal origin of the X-ray emission. Compared to NGC604, which hosts about 200 O-type stars \citep{hunt96} and is $\sim$1.5 times larger, the number of SNe expected for IC131-se is $\sim$130, which is considerably larger than the estimate from the IMF. So either the IMF estimate suffers from incompleteness and the number of O stars was significantly larger or the diffuse X-ray emission is not predominantly thermal.

The energy density of the gas amounts to $\sim$$8.0 \times 10^{-10}$\,erg cm$^{-3}$ and the corresponding cooling timescale is $\tau_{cool}\simeq3.5\, kT_e/\Lambda(T)\, n_e$, where $\Lambda(T)$ is the normalized cooling function, which amounts to $\Lambda(T)\simeq1.7\times 10^{-23}$\,erg cm$^{3}$ s$^{-1}$ for half solar metallicities \citep[][]{su93}. Hence, the cooling timescale is $\sim$1\,Gyr and much greater than the age of 8\,--\,77\,Myr of the superbubble. This implies that, due to the low density, the plasma remains hot for a long time and cannot cool efficiently. However, the main problem remains to be solved, namely how to get the plasma heated up to electron temperatures of about 4\,keV.

The total gas mass that one calculates based on a thermal model is reasonable.  The ionized gas mass traced by X-rays can be written as:
\begin{equation}
M_{\rm X}= 1.17\ m_{\rm H}\, n_{\rm e}\,V f_{\rm X}, 
\end{equation}
with $1.17\times m_{\rm H}$ being the hydrogen mass per electron, accounting for the contribution of 10\% He and assuming fully ionized gas of solar abundances. Adopting again a filling factor of the HIM of 90\%, the total X-ray gas mass amounts to $M_{\rm X}\simeq3300\pm 550$$M_{\odot}$, which is about 50\% of that in NGC604.  This is considerably greater than the mass of the progenitors of the 118$\pm$59 O-stars needed to heat the plasma, but conductive mass loss from the walls of the supershell can easily account for the mass deficit.  

Hot plasmas are expected to expand rapidly and eventually break out from their host environment \citep[e.g.,][]{tan02}. The expansion of the X-ray gas is clearly suggested by a comparison between the internal pressure of the bubble and the external pressure from the surrounding ISM.
The thermal pressure of the hot gas inside the bubble can be estimated to be $P_{\rm th}/k=1.92\,n_e T_e = 4.5\times 10^{6}$\,K\,cm$^{-3}$, assuming a filling factor of 90\%. This pressure is more than a factor of 10\,--\,20 higher than in the N11 and N51D superbubbles and almost 2 times higher than in NGC604 \citep{tull08}. Compared to the canonical value for the external pressure of $P_e/k=10^4$\,K\,cm$^{-3}$ of the ISM, a rapid expansion of the HIM is expected and even breakout seems possible given the non-detection of cooler gas which confines the HIM towards the east of the bubble.
A bubble which suffers breakout would also be consistent with an evolved age of IC131-se. If we assume that the HIM broke out from the bubble, it remains to be understood why the internal pressure is still so high. It could be that we fortuitously witness breakout at a very early stage where the pressure is still high. Assuming that this phase is short compared to the bubble's lifetime, it appears rather unlikely that we observe such a rare event. In case we do, it could explain why we have not observed similar objects before. 
Although a thermal origin is expected from the standard \hii\ region or superbubble models, electron temperatures of 4\,keV are clearly not. Even relatively young SNRs like E0102 \citep{hugh00} or 0509-67.5 \citep{chen08} have spectra which, when characterized by simple thermal models, have apparent electron temperatures significantly below 1\,keV.
It is in particular hard to understand how one could create a very hot plasma in a situation where a substantial portion of the gas mass was conducted from the walls of the region.    
Therefore, the fundamental theoretical challenge with a thermal explanation is to understand how electron temperatures of the observed order can be obtained. Shock waves in SNRs can in principle produce ion plasma temperatures between $10^7$\,--\,$10^8$K (860\,--8600\,eV) in regions behind the reverse shock. However, the electron temperature should be about three orders of magnitude lower due to mass-proportional heating, which is too low to produce X-rays. So any SNR seen in X-rays must have some amount of collisionless heating. Two such examples, for which the heating appears not to be proportional to the ion to electron mass ratio are SN\,1006 \citep[e.g.,][]{lami96} and DEM\,L71 \citep{rako09}. In these cases the Buneman instability \citep{bun58} and the excitation of hybrid waves \citep{lesch90} in non-radiative shocks are assumed to raise the electron temperature by about a factor of 300. It remains unclear whether these processes also occur in IC131-se and questionable if the additional heating is large enough to produce the observed high electron temperatures. 

If the age of IC131-se is long compared to the Coulomb equilibration timescale, the electron temperature should be high, as the protons would have had enough time to transfer their energy to the electrons via Coulomb collisions. The estimated equilibration time $t_{eq}=6.75\times T_e^{3/2}/n_e$ \citep[][]{spit78} for IC131-se amounts to 1.5\,Myr. This time is much shorter than the estimated age of 8\,--\,77\,Myr and suggests that the high temperatures are caused by Coulomb heating of the electrons. However, the Coulomb timescale in NGC604 is $\simeq$ 0.025Myr and substantially shorter than the age of the stellar population of $\sim$3\,Myr \citep{gonzo00}, yet the electron temperature there is much lower than the one predicted for IC131-se by the thermal model. Thus, it remains puzzling why the electron temperature would be so much higher in IC131-se than in other GHRs.

\subsubsection{Is the X-ray emission non-thermal?}
Based on the fits, a pure non-thermal plasma is statistically as justified as a pure thermal plasma to explain the origin of the X-ray emission in IC131-se. Non-thermal X-ray emission is observed in increasing numbers of SNRs, ranging from young objects like Cas A and SN1006 \citep[e.g.,][]{koya95,long03,hevi08} to older objects with slower shocks, such as G156.2+5.7 \citep{katsuda09}. In such objects, particles appear to be accelerated by diffusive shock acceleration \citep{bell78,malkov01} and non-thermal X-rays are most likely due to synchrotron emission or inverse Compton scattering.

Because SN1006 or 30\,Dor C show distinct cap-like regions which emit strong synchrotron X-rays, one could therefore speculate that the hard X-ray emission in IC131-se is similar in nature, except that in IC131-se the emission is projected towards the observer so that it is seen end-on rather than edge-on.  Non-thermal emission has also been invoked to explain the hard tails in the spectra of other \hii\ regions and superbubbles in the Milky Way \citep[e.g.,][]{muno06,wolk02} and LMC \citep{mad09}. IC131-se would simply be extreme in that the extent and amount of the non-thermal emission would be much larger than in any of the SNRs and \hii\ regions mentioned above.  There are at least two objects of roughly comparable size to IC131-se for which non-thermal components have been reported.  These are 30\,Dor C \citep{smith04} and N51D \citep{coop04} whose extents are $\sim$\,90 x 100 pc and 120 x 200 pc, respectively. In 30\,Dor C, \citet{smith04} found that roughly half of the total emission is thermal and half is non-thermal. The non-thermal emission is brightest along the western edge of the bubble and has a luminosity of about three times the total X-ray luminosity of IC131-se. In N51D, about 30\% of the total X-ray luminosity is non-thermal \citep{coop04} and has a luminosity of about one eighth of that observed in IC131-se. The photon index that is required to explain the X-ray emission in IC131-se of $\Gamma\simeq2.1$ is consistent with the results reported by \citet{bam04} and \citet{smith04} for 30\,Dor C.  

Recent radio observations of M33 at 20\,cm and 3.6\,cm revealed that about 7\,-\,30\% of the radio emission in IC131 is synchrotron emission \citep[0.29\,mJy and 1.33\,mJy, respectively,][]{tab07}. This radiation indicates the presence of relativistic CRs, which can also be the cause of synchrotron emission in the X-ray regime. Alternatively, non-thermal X-rays can also be produced by inverse Compton scattering of relativistic particles which interact with low energy photons from IC131 or the Cosmic Microwave Background. 
The challenge associated with a non-thermal interpretation of the emission we observe in IC131-se is that there are no detailed theoretical models for confronting the observations of IC131-se or any of the other objects in which harder X-ray components have been observed.  The basic idea is that multiple shocks from SNe and stellar winds reverberating through the bubble provide a fertile region for the acceleration of electrons and perhaps cosmic rays. It is generally believed that synchrotron radiation accounts for the non-thermal emission from SNRs.

To determine which of the non-thermal mechanisms might be the dominant one, we start with an estimate of the radiative losses due to inverse Compton scattering. The total energy lost by one electron per unit time that passes through a photon field of energy density $U_{rad}$ can be written as:
\begin{equation}\label{eq3}
\left(\frac{dE}{dt}\right)_{IC}= \frac{4}{3} \sigma_{T} c \beta^2 \gamma^2 U_{rad}\,,
\end{equation}
where $\sigma_{T}$ is the Thompson cross-section, $\beta= v/c$, and $\gamma$ is the Lorentz factor \citep[see, e.g.,][]{rl86,longair}. 
The stellar radiation field can be approximated by $U_{rad}=L_{stars}/4\pi cR^2$, where $L_{stars}$ is the luminosity of the stars and $R$ is the radius of the X-ray emitting volume in cm. Because an accurate measure of $L_{stars}$ is lacking, we assume a conservative upper limit for the 16 B-type stars inside the bubble of $4.9\times 10^{37}$\ergs, which results in $U_{rad}=1.4\times 10^{-15}$\,erg\,cm$^{-3}$. This energy density is 300 times smaller than the energy density for the cosmic microwave background of $U_{CMB}=4.2\times10^{-13}$ erg cm$^{-3}$ \citep{longair}. 

Because the relation for radiative losses from synchrotron emission is identical to Eq.~(\ref{eq3}), except that the radiation energy density $U_{rad}$ is replaced by the energy density of the magnetic field $U_{mag}=B^2/8\pi$, the ratio of the two loss rates is simply $(dE/dt)_{IC}/(dE/dt)_{syn} = U_{rad}/U_{mag}$. Assuming a magnetic field strength across IC131-se of 20$\mu$G (F. Tabatabaei, priv. comm.), the corresponding magnetic energy density is $U_{mag}=1.6\times 10^{-11}$\, erg\,cm$^{-3}$, yielding $U_{mag}\gg U_{rad}$. Hence, synchrotron losses clearly dominate over losses from inverse Compton scattering for the adopted magnetic field strength and stellar radiation field.  

If we assume that the observed X-ray emission is synchrotron radiation, we can estimate the total power of the relativistic electron population ($\beta\simeq1$). We assume further that the number density distribution of the electrons follows a piecewise power law distribution (broken power law), where each piece has the form $N(\gamma)=N_0 \gamma^{-p}$ and extrapolate one part from the radio and the other part from the X-ray regime. Here, $p$ is the power law index of the energy distribution of the radio and X-ray electrons. $p$ is related to the radio photon index $\Gamma_r$ via $p=1-2\Gamma_r$, whereas $p=1+2\Gamma_x$ holds for X-ray energies. For the radio, we adopt a photon index of $\Gamma_r=0.3$ (Tabatabaei, priv. comm.) and for the X-rays we use $\Gamma_x=2.13$ (Table~\ref{t1}).  
After replacing $U_{rad}$ with $U_{mag}$ in Eq.~(\ref{eq3}), multiplying with $N(\gamma)$, and integrating over $\gamma$, we obtain:
\begin{equation}\label{eq4}
P_{syn}= 2.66\times 10^{-14}\ U_{mag}\ N_0\left( \frac{\gamma_{max}^{3-p}-\gamma_{min}^{3-p}}{3-p}\right)\ {\rm erg\ s^{-1} cm^{-3}}.
\end{equation}
From Eq.~(\ref{eq4}), we can determine the normalization constants $N_0$ for both parts of the piecewise power law distribution by integrating over the appropriate range of $\gamma$ for the radio and X-ray emission, assuming that the electrons emit at the critical frequency $\nu_c= \gamma^2 eB/2\pi m_e$ \citep[e.g.,][]{longair}. The luminosities in these bands amount to $L_{rad}=2.8\times 10^{33}$\,erg s$^{-1}$ \citep{tab07} and $L_X=8.7\times 10^{35}$\,erg s$^{-1}$.

The total energy, $E_{tot}$, stored in the relativistic electron population can be written as:
\begin{equation}
E_{tot}=\int N(\gamma) E_e(\gamma) d\gamma dV= N_0 m_e c^2 V \left(\frac{\gamma_{max}^{2-p}- \gamma_{min}^{2-p}}{2-p}\right),
\end{equation}
where the first power law component is integrated from $\gamma_{min}=0$ to $\gamma_{break}$ and the second segment from $\gamma_{break}$ to $\gamma_{max}=\infty$. The Lorentz factor at the break point of the broken power law can be estimated by extrapolating the power law segments to where they intersect, yielding $\gamma_{break}=1.6\times 10^6$. 
The total energy in the synchrotron emitting electron population is therefore $2.83\times 10^{51}$\,erg which is $\sim$14\%\,--30\% of the total energy provided by the estimated number of SNe (9\,--\,21). It follows that non-thermal synchrotron emission can be a significant contributor to the overall emission.
This result is somewhat unexpected and the challenge to this interpretation is to determine how a relatively strong magnetic field can be maintained on large scales in the absence of colliding winds from O stars. This result is also in contrast to \cite{muno06} who found that inverse Compton scattering is the most likely loss mechanism to explain the non-thermal X-ray emission in Westerlund\,1.

\subsubsection{Conclusions }
From an observational perspective, we have shown that IC131-se is an exceptional object in terms of the combination of the extent and the hardness of its X-ray spectrum. 
We have explored the feasibility of various emission mechanisms, and we conclude that, with the available data, we cannot distinguish between interpretations based on thermal emission from a hot ISM filling the interior of the superbubble or non-thermal radiation which seems to originate from synchrotron photons. The primary challenge set down by the presented data is for others to attempt to produce models of superbubbles and GHRs that yield high enough electron temperatures to reproduce the spectrum or enough high energy particles to generate the X-rays that are observed via synchrotron radiation. It would be interesting to know what makes IC131-se so peculiar, and if there are any other GHRs which are really similar. 

\section{Point source CXO\,J013315.10+304453.0}\label{point_dis}
\subsection{Imaging results}
\begin{figure}[t]  
\centering
\includegraphics[width=8.25cm,height=6.5cm,angle=0]{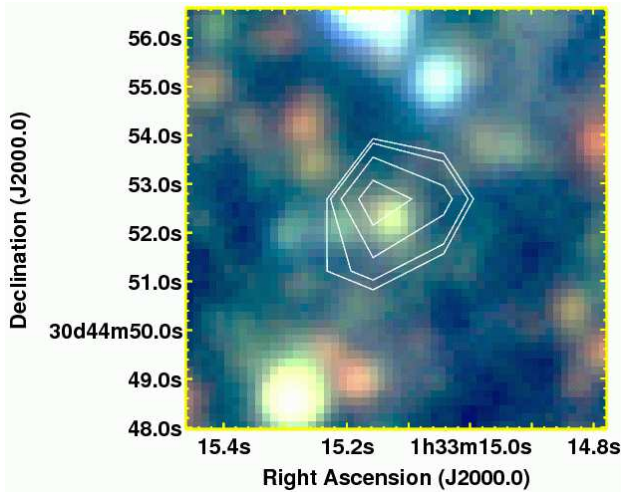}
\caption{\label{f5}
A zoom into the region highlighted in Fig.~\ref{f1}f revealing a possible optical counterpart to FL073. Contours are from the unbinned merged event list in the 0.35\,-8.0\,keV energy band and are plotted on a square root scale between 2 and 8 counts per pixel. See the electronic edition of the Journal for a color version of this figure.}
\end{figure}
The distribution of the diffuse X-ray emission in IC131-se in the soft, medium, and hard energy band (Fig.~\ref{f1}e) reveals several localized enhancements of the X-ray emission, some of which can be associated with optical counterparts.
Based on the full \chase\ source catalog (which will be presented elsewhere), the only X-ray point source in IC131-se, whose extent is consistent with the PSF, is CXO\,J013315.10+304453.0 (FL073). 

This object is visible in Fig.~\ref{f1}e as a white circular region (see red plus sign) and seems to have an optical counterpart (Fig.~\ref{f5}) which can be identified as J013315.17+304452.5, a faint star of visual magnitude $m_{\rm V}=22.2$ and $B-V=0.14$ \citep{mass06}. 
Using $E(B-V)=0.07$ \citep{mass06} and a distance of 817\,kpc \citep{freed01} the apparent visual magnitude translates into an absolute magnitude of $M_{V}$\,=\,$-2.61$. Although optical spectroscopic data, which could clarify the spectral type and a possible membership to IC131-se is missing, color and visual magnitude seem to imply a star of spectral type B2{\small V} with a mass of about 9M$_{\odot}$  \citep{sk82}. Such high-mass stars are often found as companion stars in high-mass X-ray binary (HMXRB) systems \citep[e.g.,][]{mcb08}. 

\begin{deluxetable}{l|cccc}
\tablecaption{\label{t2} Fit parameters (Point source)}
\startdata
\hline\hline \\[-0.5cm]
tbabs $\times$ (tbvarabs\ldots)\tablenotemark{a}               & (pow)                 & (brems)              & (apec)               & (nei)               \\[0.1cm]
\hline \\[-0.5cm]
$\chi^{2}_{\rm red}$                                            & 1.09                  & 1.13                 & 1.11                 & 1.10                \\       
degrees of freedom                                            & 15                    & 15                   & 15                   & 14                   \\
$P$ value                                                     & 0.53                  & 0.58                 & 0.59                 & 0.52                 \\[0.1cm] 
$K$ ($10^{-6}\,{\rm cm}^{-5}$)\tablenotemark{b}                & 1.20$^{+0.48}_{-0.23}$  & 1.49$^{+0.42}_{-0.28}$ & 4.46$^{+0.79}_{-0.74}$ & 4.40$^{+0.73}_{-0.75}$ \\[0.1cm]
\hline \\[-0.5cm]
$N_{\rm H}$ ($10^{22}\,{\rm cm}^{-2}$)                          & $<$0.01$^{+0.26}_{-0.01}$  & $<$0.01$^{+0.17}_{-0.01}$ & $<$0.01$^{+0.17}_{-0.01}$ & $<$0.01$^{+0.16}_{-0.01}$ \\
$\Gamma$\tablenotemark{c}                                     & 1.88$^{+0.42}_{-0.26}$  & ---                  & ---                  & ---                  \\ 
$kT_e$ (keV)\tablenotemark{d}                                 & ---                   & 3.89$^{+4.07}_{-1.62}$ & 3.96$^{+3.77}_{-1.42}$ & 3.98$^{+3.68}_{-1.47}$ \\
$\tau$ ($10^{13}$\,s/cm$^3$)\tablenotemark{e}                  & ---                   & ---                  & ---                  & 4.76$^{+4.76}_{-4.76}$ \\
$\tilde{F}_{\rm X}$ ($10^{-15}$\,erg/s/cm$^2$)\tablenotemark{f} & 5.87$^{+0.66}_{-2.03}$  & 5.26$^{+0.71}_{-3.16}$ & 5.36$^{+0.57}_{-2.00}$ & 5.37$^{+0.92}_{-2.52}$ \\
$F_{\rm X}$ ($10^{-15}$\,erg/s/cm$^2$)\tablenotemark{f}         & 6.41$^{+0.72}_{-2.22}$  & 5.66$^{+0.90}_{-4.00}$ & 5.74$^{+0.61}_{-2.14}$ & 5.74$^{+0.98}_{-2.69}$ \\
$L_{\rm 0.35-8}$ ($10^{35}$\,erg/s)\tablenotemark{g}            & 1.28$^{+0.14}_{-0.44}$  & 1.13$^{+0.18}_{-0.80}$ & 1.14$^{+0.12}_{-0.43}$ & 1.14$^{+0.20}_{-0.54}$ \\[0.1cm]
$L_{\rm 2-8}$ ($10^{35}$\,erg/s)\tablenotemark{h}              & 0.62$^{+0.06}_{-0.18}$  & 0.55$^{+0.07}_{-0.30}$ & 0.58$^{+0.05}_{-0.17}$ & 0.58$^{+0.08}_{-0.21}$  \\[0.1cm]
\enddata
\vspace{-0.5cm}
\tablenotetext{a}{See Table~\ref{t1} for a description of the parameters.}
\end{deluxetable}
\begin{figure*}
\centering
\includegraphics[width=16.5cm,height=6cm,angle=0]{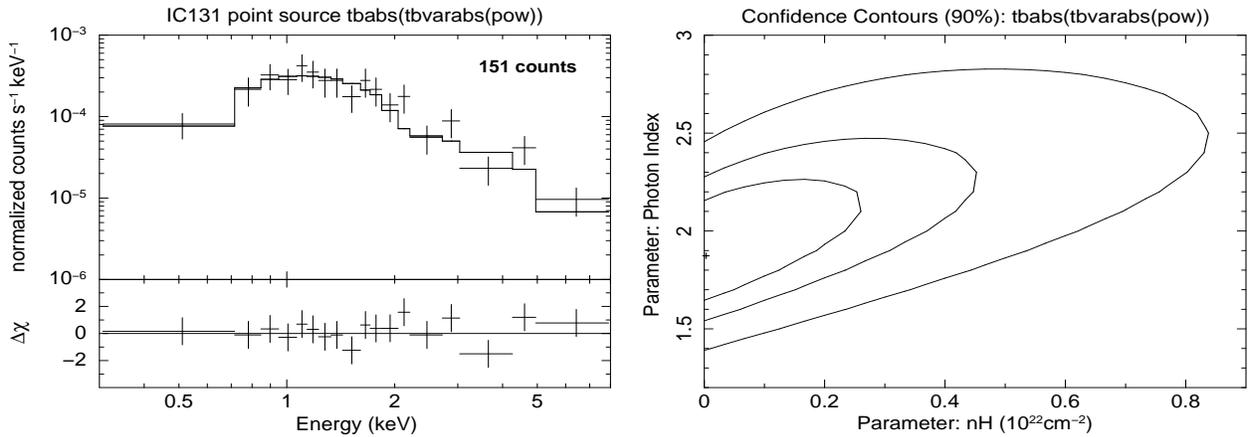}
\caption{\label{f6}
Shown are the spectrum of CXO\,J013315.10+304453.0 (FL073) and the best model fit of the absorbed power law (left panel) together with the 90\% confidence contours (right panel).}
\end{figure*}
\subsection{Spectroscopic Results}
Because the point source has only $\sim$150 counts, we used only absorbed single component models to fit the X-ray spectrum of FL073. The \hi\ absorption was again modeled with a combination of {\tt tbabs} and {\tt tbvarabs} models, assuming the same values for the Galactic and internal absorption in \m\ as in Section \ref{spec_res}. In Table~\ref{t2} the best-fit results are shown for a non-thermal and a number of thermal plasma models. Fig.~\ref{f6} presents the spectrum and 90$\%$ confidence contours of the adopted power law model. Because the point source appears to be embedded in diffuse gas, the background was taken from regions of the diffuse emission. As a consequence, a relatively hard background spectrum was subtracted which apparently softened the net emission and required only a marginal internal column density. However, the upper limit on the absorption is in good agreement with the total integrated column density. 

Unfortunately, the fit statistics and unabsorbed luminosities are very similar for all models and do not favor one model over another. Because there are no indications of emission line features and the spectrum looks relatively smooth, we assume the absorbed power law model to be the best fit to the data. The unabsorbed broadband X-ray luminosity of FL073 amounts to about $1.3\times 10^{35}$\ergs, $\sim$50\% of which are emitted at hard X-rays above 2.0\,keV. 

\subsection{Discussion}
The photon index of FL073 of $\Gamma=1.88^{+0.42}_{-0.26}$ is consistent with that of two transient X-ray sources in \m, called XRT-2 or XRT-5 in \citet{ben08} which have photon indices of 1.6$\pm 0.4$ and 1.9$\pm0.5$, respectively. XRT-2 seems to have an optical counterpart which could be an early B star with very similar color ($B-V=0.1$) and visual magnitude ($m_{V}=22.8$) compared to the counterpart of FL073. The XRB hypothesis is corroborated by FL073's hardness ratios of $HR1=-0.17$ and $HR2=-0.13$ \citep[see][for definition and adopted energy bands]{plu08}. They agree well with those of the other XRB candidates that can be fitted with an absorbed power law $<2.0$ and an \nh\ of $\sim 10^{21}$\,cm$^{-2}$ (see their Fig.~5). Although the photon index and the hardness ratios for FL073 could also indicate an AGN, the column density suggests that this object is part of IC131-se and is likely not a background AGN.

\begin{figure}[t]
\centering
\includegraphics[width=7.6cm,height=5.5cm,angle=0]{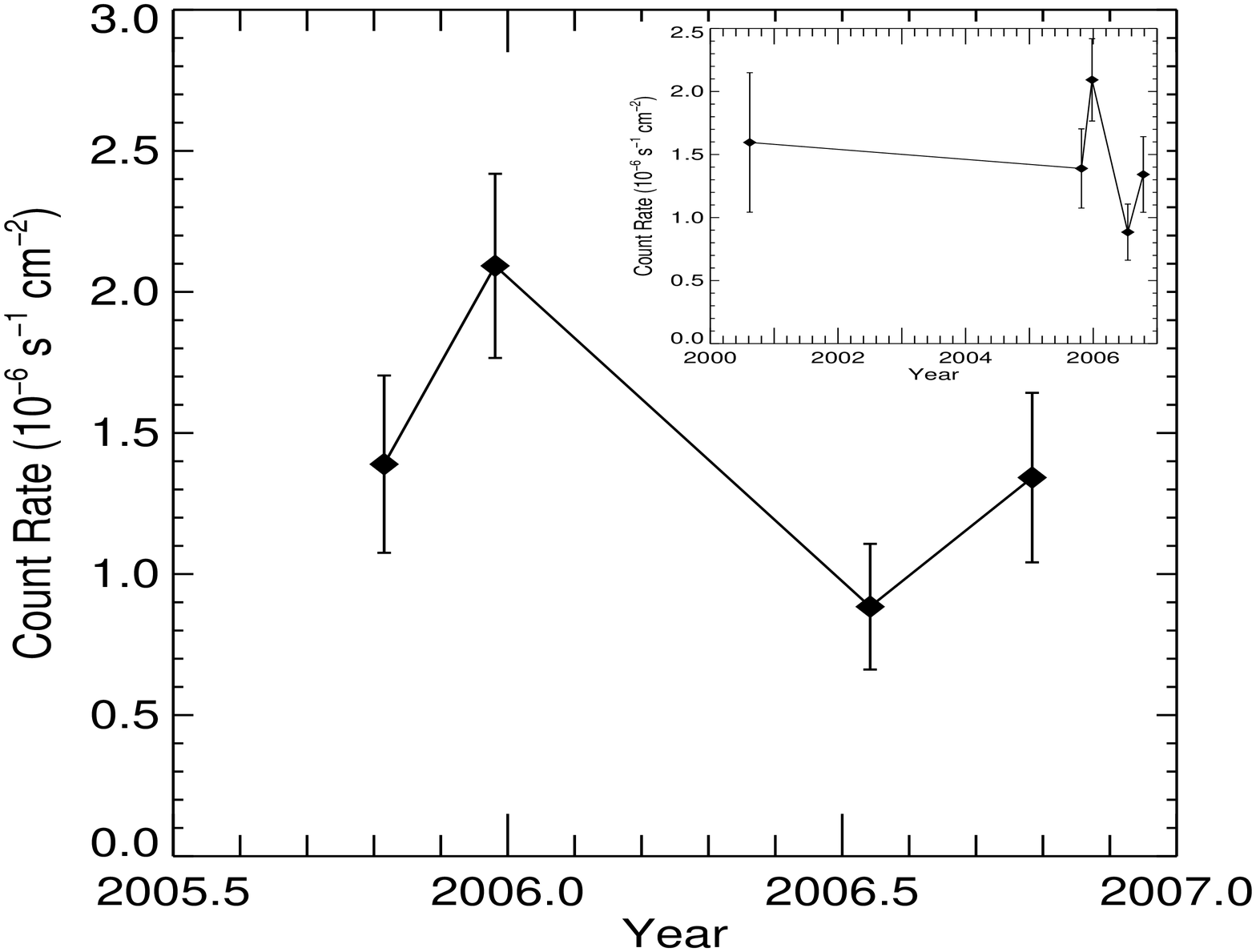}
\caption{\label{f7}
Long-term variability of FL073 (the inlay shows the full lightcurve, including the archival ObsID 1730).}
\end{figure}
Based on the data presented by the \rosat\ \citep{haberl01} and \xmm\ surveys of \m\ \citep{pietsch04,misa06} it cannot conclusively be determined whether FL073 was detected, as the poor spatial resolution of these satellites did not allow for a clear separation of point sources and diffuse emission. 
FL073 is not listed in the \m\ CFHT optical variability survey \citep{hart06} as a variable source. In X-rays this source is detectable in all ObsIDs and reveals a possible long-term variability (Fig.~\ref{f7}). Fitting the lightcurve with an assumed constant flux yields $\chi^2/\nu=9.47/3$, which corresponds to a 3\% probability of being constant. Although the amplitude of the variation is not strong ($\sim 2.2\sigma$), we suggest that FL073 is associated with a HMXRB system in which the optical companion is a B2-type star with a mass of $\sim$9M$_{\odot}$. It should be noted that only the high angular resolution of \cxo\ allowed a proper treatment of this source whose luminosity is $\sim$16\% of the total diffuse X-ray luminosity. At lower resolution this point source would have seriously contaminated the emission from IC131.

\section{Summary and conclusions}\label{con}
Compared to known superbubbles and \hii\ regions IC131-se appears to be peculiar with respect to its lopsided X-ray morphology, the large linear extent of the X-ray emission, the lack of massive O stars, its high electron temperature (in case the X-ray gas is thermal), and large fraction of hard X-rays. 
The X-ray spectrum of the extended emission in IC131-se can be equally well fitted by an absorbed power law ($\Gamma \simeq 2.1$) or an absorbed thermal plasma model ($kT_e\simeq 4.3$\,keV). These models predict a total (0.35\,--\,8.0\,keV) unabsorbed X-ray luminosity of about $9\times10^{35}$\ergs\ and $7\times10^{35}$\ergs, respectively, with 39\% and 53\% of the luminosity being emitted above 2\,keV. Apparently IC131 possesses not only the hardest X-ray spectrum among the known GHRs, but, in case the gas is thermal, also the hottest X-ray plasma.

Besides the facts that thermal X-ray emission is expected from GHRs and that there might be weak emission lines below 1\,keV, the disagreement between the number of SNe progenitors obtained from the IMF (9\,--\,21) and that derived from the internal thermal energy (118$\pm$59) argues against a pure thermal origin of the gas. We estimate a low electron density of $n_e\simeq 0.05$\,cm$^{-3}$ (assuming $f_X =0.9$) and an X-ray gas mass of about 3300\,$M_{\odot}$. The age of the stellar population inside the bubble seems to range from 8 to 77\,Myr and given a cooling timescale of about 1\,Gyr, the gas did not have time to cool significantly.
If the gas is thermal, the standard bubble models require shocks from SNRs and SNe to heat the gas. 
The fundamental problem which argues against a thermal plasma is that we are not aware of a mechanism for GHRs which can produce electron temperatures anywhere close to $kT_e=4$\,keV. 

Non-thermal X-ray emission is also known to originate from SNRs, GHRs or superbubbles and the spectral power law fit is statistically as good as the one for a thermal plasma. Synchrotron emission as opposed to inverse Compton scattering is a plausible mechanism for the assumed magnetic field strength of $B=20\mu G$ and the total energy of the relativistic synchrotron-emitting electron population is about 14\% to 30\% of the total energy provided by SNe. If a non-thermal component exists, synchrotron losses clearly dominate over losses from inverse Compton scattering.
For a purely non-thermal origin, however, it remains to be investigated what kinds of mechanisms are able to produce non-thermal X-rays or accelerate particles to relativistic energies on scales of 200\,pc. To allow for a more quantitative analysis of the non-thermal X-ray emission, models are required which predict the fraction of non-thermal radiation from GHRs.
A combined thermal and non-thermal model seems to be generally appropriate, too, irrespective of the unreasonably high column density and X-ray luminosity derived from those fits. 

Unfortunately, with the present data the nature of the extended X-ray emission cannot be conclusively determined. Clearly, IC131-se is an important object and challenges the standard bubble model as well as our understanding of CR acceleration in superbubbles. It is remarkable and remains to be understood why objects similar to IC131-se have not been detected before. Future investigations would greatly profit from more sensitive and deeper X-ray observations (as for example provided by IXO), high-resolution optical spectrophotometric data, and models which can explain electron temperatures $>$\,1\,keV and make predictions on the non-thermal X-ray emission in GHRs. The deeper X-ray observations could establish the presence of X-ray emission lines and settle the question on the nature of the emission, while a detailed stellar population analysis could provide a more accurate IMF. Clearly, NGC604 and IC131-se seem to be in completely different evolutionary stages. All O-type stars in IC131-se seem to have exploded as SNe, whereas the western part of NGC604 awaits the first SNe to occur. 

We detect only one X-ray point source in IC131. This source, CXO\,J013315.10+304453.0 (FL073), appears to be time variable and is possibly a HMXRB with an optical counterpart which could be a B2V star with a mass of about 9M$_{\odot}$.

\acknowledgments
This work has made use of SAOImage DS9 \citep{joy03}, developed by the SAO, the FUNTOOLS utilities, and the HEASARC FTOOLS package. We thank the anonymous referee for useful comments which helped to improve the paper. RT acknowledges support under NASA \cxo\ award number GO6-7073A. P. F.W. acknowledges support through G06-7073C. TJG and PPP acknowledge support under NASA contract NAS8-03060.

\end{document}